\documentclass[epj]{svjour}

\usepackage{graphicx}
\usepackage{fancyhdr}
\usepackage{textcomp}
\def\makeheadbox{{
 }}
\def\mytitle{Determination of the Dark Matter profile} 
\def\myauthors{Markus Weber}    
\def\mytype{Parallel}    
\def\mysession{Cosmology and Astrophysics}

\begin{document}
\title{Determination of the Dark Matter profile from the EGRET excess of diffuse Galactic gamma radiation}
\author{Markus Weber\inst{\mail{Markus.Weber@ekp.uni-karlsruhe.de}}
}
%
%
\institute{Institut f\"ur Experimentelle Kernphysik, Universit\"at Karlsruhe (TH), P.O. Box 6980, 76128 Karlsruhe, Germany
}
%
\date{}

\abstract{The excess above 1 GeV in the energy spectrum of the diffuse Galactic gamma radiation, 
measured with the EGRET experiment, can be interpreted as the annihilation of Dark Matter (DM) particles. 
The DM is distributed in a halo around the Milky Way. Considering the directionality of the gamma ray flux it is possible to determine the halo profile. The DM within the halo has a smooth and a clumpy component.
These components can have different profiles as suggested by N-body simulations and the data is indeed compatible with a NFW profile for the diffuse component and a cored profile for the clumpy component.
These DM clumps can be partly destroyed by tidal forces from interactions with stars and the gravitational potential of the Galactic disc.
This effect mainly decreases the annihilation signal from the Galactic centre (GC). In this paper constraints on the different profiles and the survival probability of the clumps are discussed.
} 
\maketitle
%
%
\section{Introduction}
\label{seq:intro}
From WMAP measurements of the temperature aniso- tropies in the Cosmic Microwave Background (CMB) in combination with data on the Hubble expansion and the density fluctuation in the Universe \cite{RefSpergel} we gather that Cold Dark Matter (CDM) makes up 23\% of the energy of the Universe. 
The nature of the Dark Matter (DM) is unknown, but one of the most promising particles is the "weakly interacting massive particle" (WIMP).
Assuming that WIMPS are Majorana particles they can annihilate each other and produce a large amount of secondary particles.
For the determination of the density distribution, the so-called DM halo profile of the WIMP particles, the gamma radia\-tion is most important because it is not influenced by the magnetic field of the galaxy and points back directly to its source.\\
The observation of the diffuse Galactic gamma radiation of the Milky Way with EGRET showed an excess above 1 GeV in the photon energy spectrum. This excess is different for various sky directions and can be interpreted as a WIMP annihilation signal \cite{RefSander}. Therefore it can be used to determine the DM halo profile.\\
In section \ref{seq:halo}, we will describe the mechanism of the determination of the DM halo profile from the EGRET excess and explain how to calculate the DM annihilation flux of gamma rays. Then, after differentiating between diffuse and clumpy DM, we will introduce a survival probability for DM clumps as well as a ringlike substructure of DM within the Galactic disc.

\section{Determination of the halo profile}
\label{seq:halo}
This directionality of the gamma ray excess depends on the shape of the DM halo. Therefore a fit to the signal in all different sky directions can be used to determine a possible halo profile.\\
We divided the sky into four different latitude ranges (absolute values of latitude: 0-5\textdegree, 5-10\textdegree, 10-20\textdegree and 20-90\textdegree) and 45 longitude bins. Consequently, we obtained 180 different sky regions. Strong point sources in the galaxy like CRAB and VELA were excluded because of their large contribution to the flux of the photons with a peculiar spectrum.
The local annhilation rate is proportional to the square of the number density of DM particles.
Therefore the gamma ray flux is calculated with the line-of-sight integral
\begin{equation}
  \Phi_{\chi} (E,\psi) = \frac{\langle\sigma v\rangle}{4 \pi} \sum_f \frac{dN_f}{dE} b_f \int_{los} B_l \frac{1}{2} \frac{\langle\rho_\chi\rangle^2}{M^2_\chi} dl_\psi .
\label{eq:diffflux}
\end{equation}
Here $b_f$ is the branching ratio into the tree-level annihilation final state while $dN_f/dE$ is the differential photon yield for the final state f. The value of the WIMP mass $M_\chi$ is an essential parameter for the calculation of the gamma flux and can be obtained by fitting the shape of the photon energy spectrum in different sky regions. A WIMP mass in the range of 50 - 70 GeV yields a good fit (see Ref. \cite{RefDeBoer}).
Furthermore, $\rho_\chi$ is the DM density profile of the halo and $B_l$ is a local enhancement factor, the so-called "boost factor" which characterizes the increased WIMP annihilation rate in subhalos (DM clumps) in the DM halo of the Milky Way. These DM clumps are predicted by ana\-lytical calculations \cite{RefDokuchaev3} and numerical simulations \cite{RefMoore3} from the inflationary-produced adiabatic density fluctuations. In Ref. \cite{RefSander} Eq. \ref{eq:diffflux} was used to determine the halo profile. However, for a clumpy DM distribution one expects the flux to be proportional to $\rho_{\chi,clump}$ instead of $\rho^2_{\chi,clump}$ because the clumps have a certain luminosity $\bar \rho_{cl,P}$. Therefore the flux is proportional to $\bar \rho_{cl,P} \cdot \rho_{\chi,clump}$ instead of $\rho^2_{\chi,clump}$. The boost factor can in this case be written as \cite{RefDokuchaev}
\begin{equation}
  B_l =  1 + \frac{\int \rho_{\chi,clump}(r)\ dr\ \xi_P (r)\ \bar \rho_{cl,P}(r, n_p)}{\int \rho^2_{\chi}(r)\ dr}.
\label{eq:boost}
\end{equation}
\begin{figure}
\begin{center}
\includegraphics[width=0.5\textwidth,height=0.42\textwidth,angle=0]{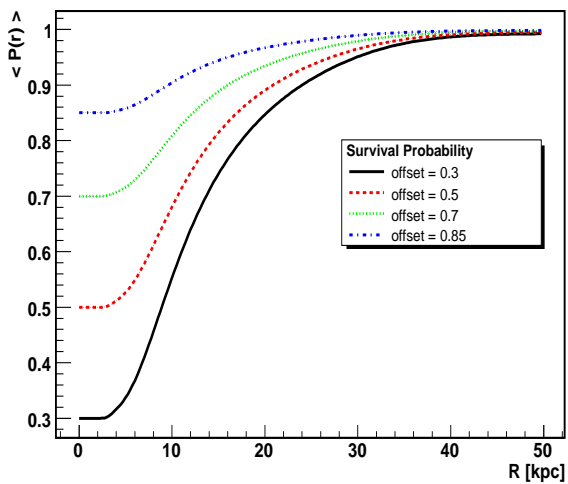}
\caption{Radial dependence of the survival probability for DM clumps}
\end{center}
\label{fig:prob}       
\end{figure}
In Eq. \ref{eq:boost} the parameter $\xi_P (r)$ characterizes the mass fraction of clumps in the DM halo, $n_p$ is the power index of primordial density perturbations.\\
The first term in Eq. \ref{eq:boost} represents the boost factor for the diffuse DM component while the second term is the boost factor for the clumpy DM.
Assuming that the mass fraction in clumps is larger than a few percent the whole annihilation rate is dominated by the clumpy DM flux which can be approximated as 
\begin{equation}
  \Phi_{DM,clump} \propto \int \xi_P (l_\psi)\ \bar \rho_{cl,P}(l_\psi, n_p)\ \rho_{\chi,clump}(l_\psi)\ dl_{\psi}
\label{eq:clumpflux1}
\end{equation}
However, the expression $\xi_P (r, n_p) \cdot \bar \rho_{cl,P}(r, n_p)$ is equal to a boost factor $B$, which is constant for all directions, multiplied with a factor $P(r)$, which is a "survival probability", averaged over all clump masses as calculated in Ref. \cite{RefDokuchaev}.
\begin{figure}
\begin{center}
\includegraphics[width=0.5\textwidth,height=0.42\textwidth,angle=0]{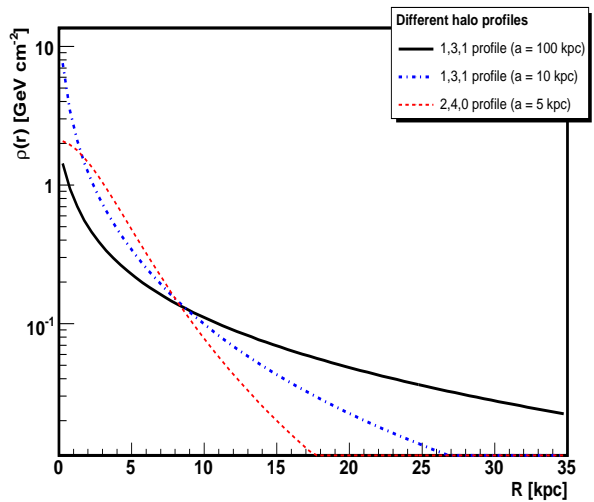}
\caption{Comparison of different halo profiles. The normalization $\rho_0$ is the same for all profiles.}
\end{center}
\label{fig:profile}       
\end{figure}
\begin{figure*}
\includegraphics[width=0.43\textwidth,height=0.45\textwidth,angle=0]{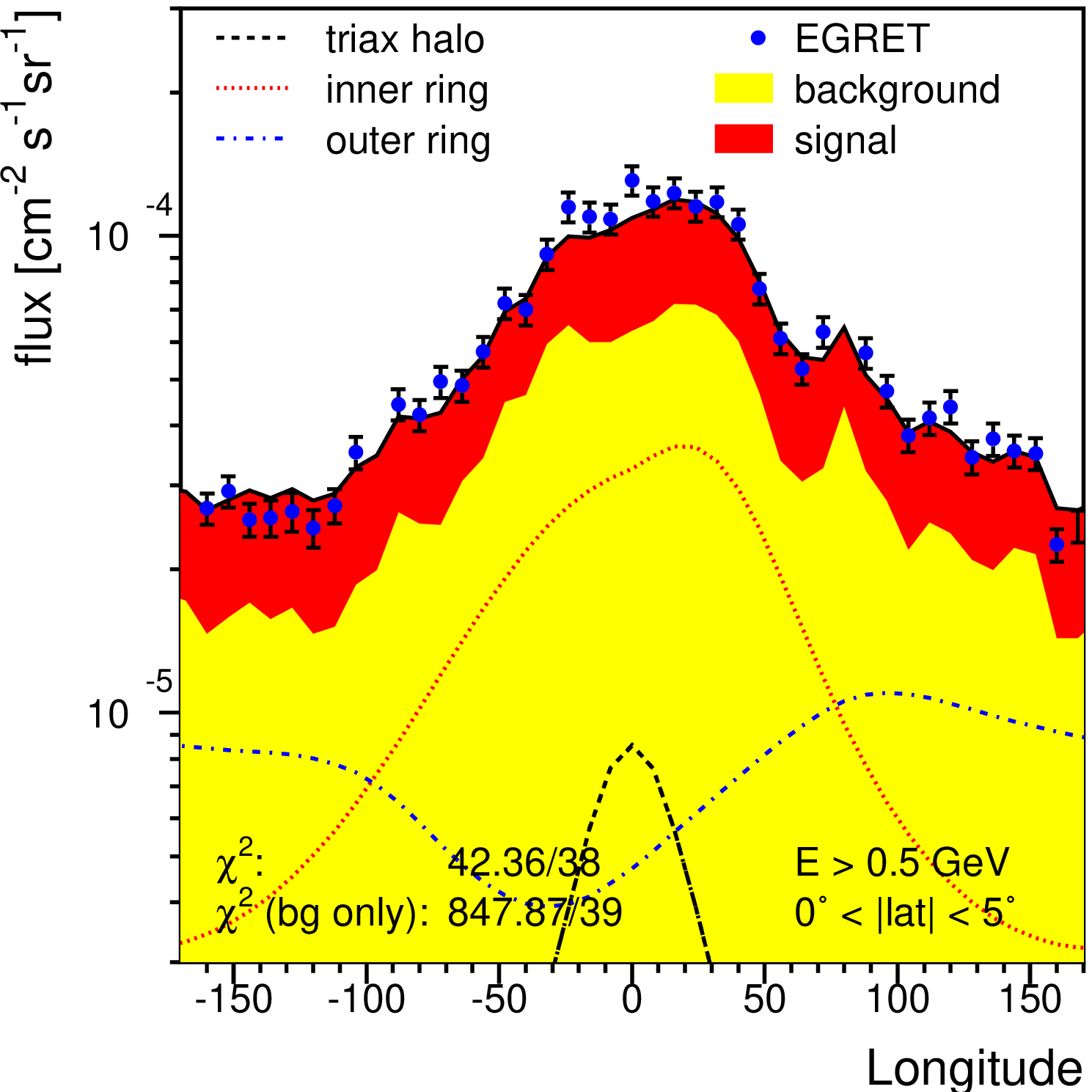}
\hfill
\includegraphics[width=0.43\textwidth,height=0.45\textwidth,angle=0]{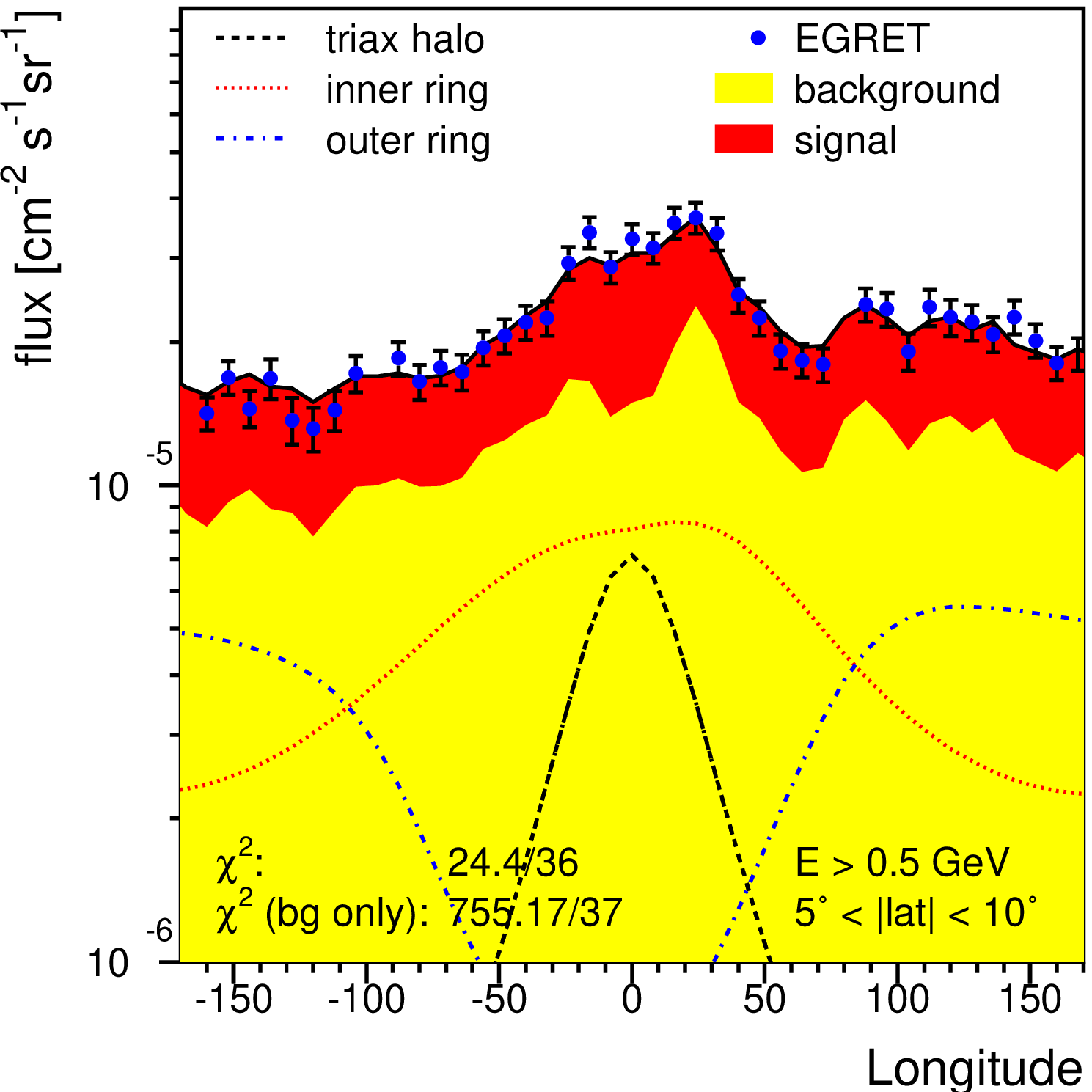}
\includegraphics[width=0.43\textwidth,height=0.45\textwidth,angle=0]{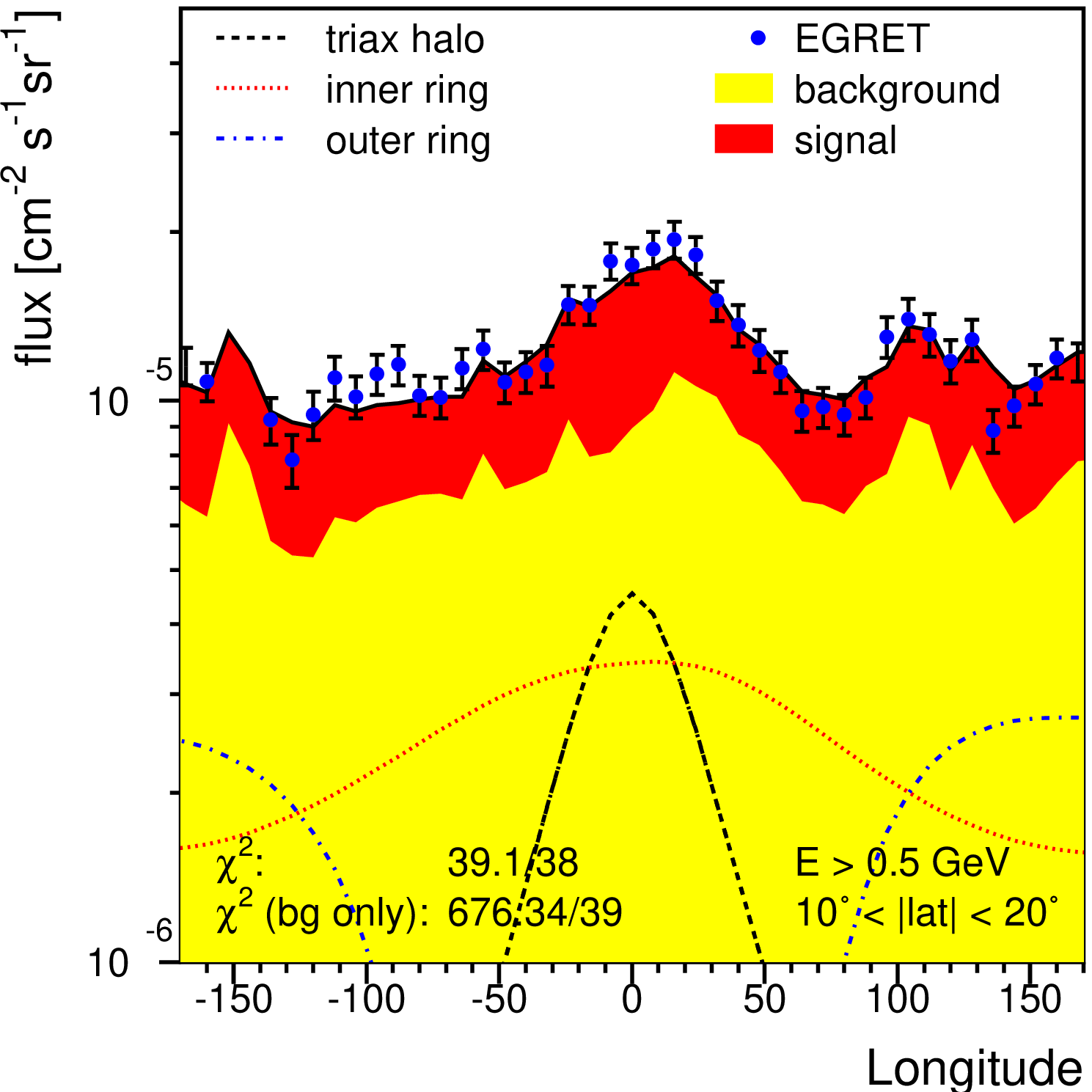}
\hfill
\includegraphics[width=0.43\textwidth,height=0.45\textwidth,angle=0]{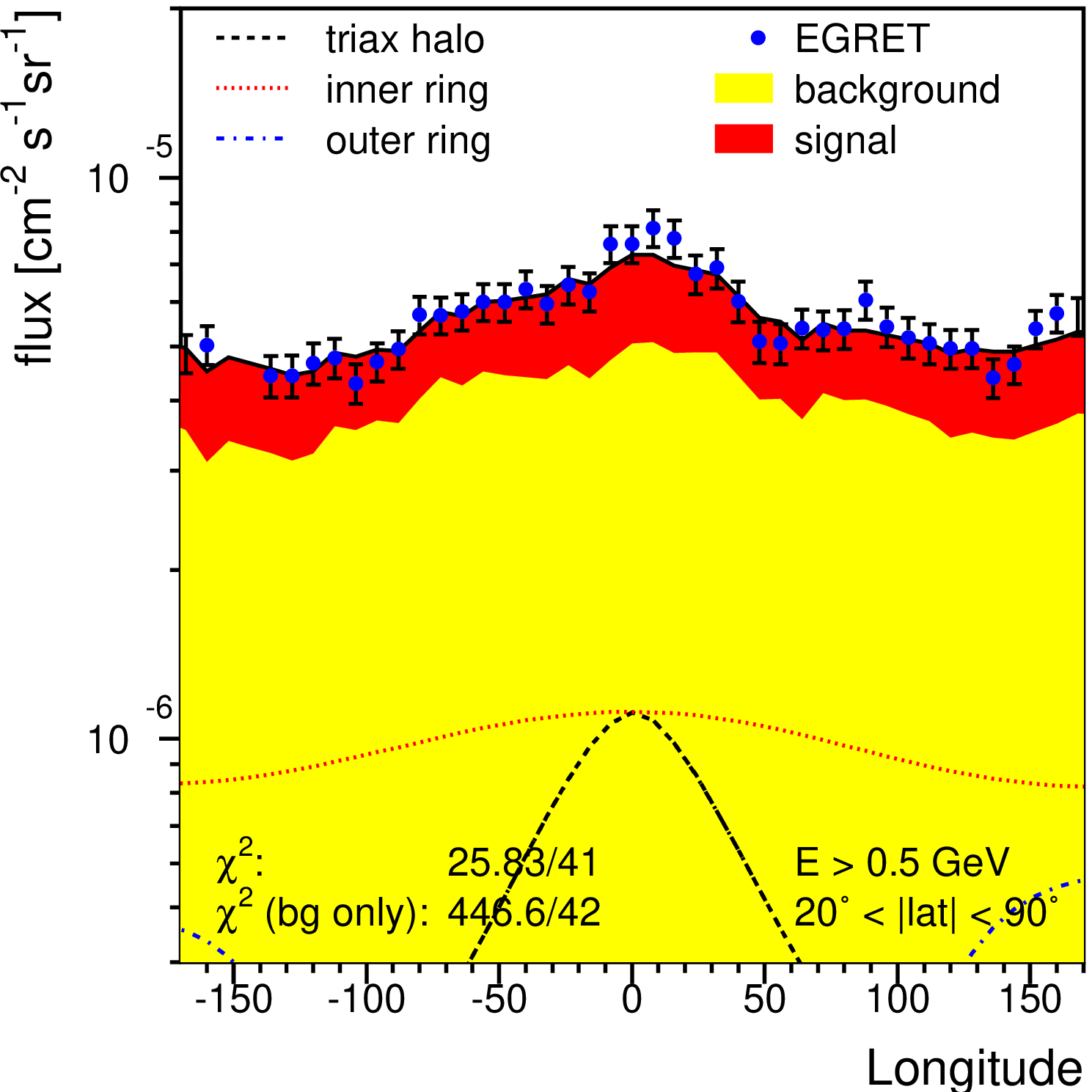}
\caption{Longitude distribution of diffuse Galactic gamma radiation above 0.5 GeV with ringlike substructures at 4 kpc and 13 kpc}
\label{fig:longitudes}       
\end{figure*}
There it was assumed that clumps are completely disrupted by tidal forces from interactions with stars and the disc potential. Therefore P(r) tends to zero for light clumps and small distances from the Galactic centre. However, more detailed calculations show that cuspy centres of the clumps survive the tidal disruption. In this case P(r) drops to a constant value of $P(0)$ near the centre of the galaxy \cite{RefDokuchaev2}. However, $P(0)$ is strongly model-dependent, especially it is sensitive to the mass distribution of the clumps and the steepness of the cusp in the centre of the clump. Therefore we kept $P(0)$ as a free parameter and left the shape of $P(r)$ roughly the same. $P(r)$ is shown in Fig. \ref{fig:prob} for various values of $P(0)$.
The flux can be written as
\begin{equation}
  \Phi_{DM,clump} \propto B \cdot \int P(r) \cdot \rho_{\chi,clump}(l_\psi)\ dl_{\psi}
\label{eq:clumpflux2}
\end{equation}
where the DM halo profile can be parametrized as
\begin{equation}
  \rho_{\chi}(r) = \rho_0 \cdot \left ( \frac{r}{r_0} \right )^{\gamma} \left [ \frac{1 + (\frac{r}{a})^{\alpha}}{1 + (\frac{r_0}{a})^{\alpha}} \right ]^{\frac{\gamma-\beta}{\alpha}}
\label{eq:haloprof}
\end{equation}
In Eq. \ref{eq:haloprof} $a$ is the scale radius of the profile and $\alpha$, $\beta$ and $\gamma$ are the radial dependencies at $r \approx a$, $r \gg a$ and $r \ll a$. The radius $r_0$ is the location of the sun and $\rho_0$ represents the DM density in the solar system, so $\rho(r_0) = \rho_0$.\\
The observation of the rotation curves of different dwarf galaxies show a flat rotation curve and a constant DM density in the Galactic centre, a so-called core. However, numerical simulations of galaxy formation prefer a Navarro-Frenk-White (NFW) profile \cite{RefNavarro} which has a cusp in the Galactic centre and decreases with $1/r^3$ for large radii ($\alpha = 1$, $\beta = 3$ and $\gamma = 1$).
Other N-body simulations predict that, in addition to these two profiles, similar profiles which are even cuspier in the centre \cite{RefMoore} are possible.\\
The question is whether the density profiles $\rho_{\chi, clump}(r)$ for the clumpy DM component and $\rho_{\chi}(r)$ for the diffuse DM component are the same or not. According to Moore et al. \cite{RefMoore2} $\rho_{\chi}$ is expected to be a cuspy profile, while $\rho_{\chi, clump}$ could be a cored profile.\\
It was shown in Ref. \cite{RefSander} that the DM annihilation signal of a spherical halo profile is not sufficient to reproduce the Galactic gamma radiation 
measured with EGRET. To explain the data two ringlike substructures in the Galactic disc were needed. A first indication of such a substructure was found by Hunter et al. in 1997 \cite{RefHunter}.
The first and inner ring is located at a radius of 4 kpc. It coincides with a ring of molecular hydrogen which is located at 4.5 kpc and has a width of approximately 2 kpc. The second ring is located at a radius of 13 kpc and most likely resulted from the infall of a satellite galaxy. This assumption was strengthened by the results of Penarubbia et al. \cite{RefPenarubbia} who found out that the tidal streams of the dwarf galaxy Canis Major form ringlike structures at about 13 kpc.\\
In the z-direction both rings are exponentially distributed with a width of 300 pc (830 pc) for the inner ring (outer ring). However the distribution in r-direction is different for the both rings. The inner ring has a Gaussian distribution in r-direction with a width of 4.3 kpc, while the outer ring has a Gaussian distribution with a width of 4.5 kpc to the outside and falling of to zero at a radius of 9 kpc by a parametrization of an s-shape with two parabolic functions.\\
Additional evidence for a ringlike substructure was found by Kalberla et al. \cite{RefKalberla}. They observed the gas flaring of molecular hydrogen in the Galactic disc and found that they need a ringlike substructure at 13 kpc to explain their measured data. This is a very impressive hint because it is not related to the EGRET observations.\\
In our analysis a good fit was obtained for a cored profile ($\alpha=2$, $\beta = 4$, $\gamma=0$, $a = 5\ kpc$) for the clumpy DM and a NFW profile ($\alpha=1$, $\beta=3$, $\gamma=1$, $a = 100\ kpc$) for the diffuse DM. The normalization parameter $\rho_0$ was found to be roughly the same for both profiles. In the case of the diffuse DM profile a large scale radius was obtained, since the Milky Way is located in a local cluster of galaxies which yields large amounts of matter at large galactocentric distances. The total DM density distributions is the summation of all DM components and can be written as
\begin{equation}
  \rho_{\chi, total}\ =\ \rho_{\chi} + \rho_{\chi,clump} + \rho_{inner Ring} + \rho_{outer Ring}
\label{eq:rho}
\end{equation}
The total mass of the galaxy is constrained by the fact that the rotation velocity at the location of the sun is $v_{sun} = 220\ km/s$. The mass of the galaxy within the virial radius $R_{200}$ is about $10^{12}$ solar masses which is in good agreement with Ref. \cite{RefBattaglia}. Note that the diffuse component hardly contributes to the DM annihilation signal, but significantly to the mass. The density profiles are compared in Fig. \ref{fig:profile}.\\
The parameters of the halo profiles and the boost factor $B$ are varied to minimize the $\chi^2$ function which is described in Ref. \cite{RefSander}. The constant boost factor $B$ is found to be about 90 and $P(0)=0.7$ which agrees well with the data and is in the expected range \cite{RefDokuchaev2}.\\
The best fit of the longitude distribution is shown in Fig. \ref{fig:longitudes} and the DM density profile with rings is shown in Fig. 4.\\


\begin{figure}
\begin{center}
\includegraphics[width=0.43\textwidth,height=0.43\textwidth,angle=0]{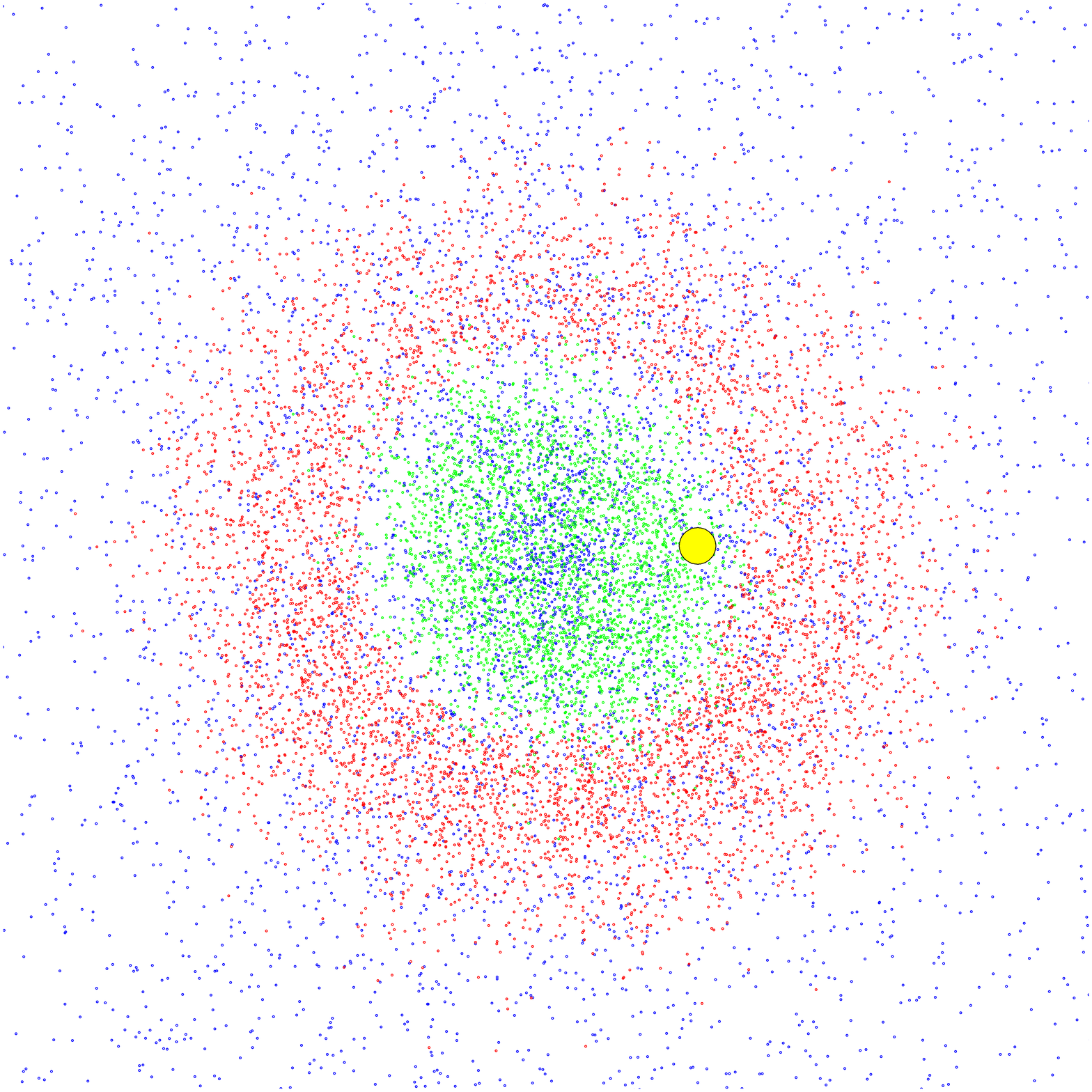}
\caption{Density of the DM in the Galactic plane. The sum of the diffuse and clumpy DM are shown in blue dots, while the inner and outer ring are illustrated in green and red dots. The position of the sun is marked with a yellow point}
\end{center}
\label{fig:scatter}
\end{figure}
\section{Summary}
CDM probably consists of WIMPS which are heavy and slow particles. If these particles are Majorana particles they can annihilate each other and produce Galactic gamma radiation which can be used to determine the density profile of the DM. In this analysis the directionality of the DM annihilation flux measured with EGRET was used to find a possible DM halo profile. After dividing the Galactic gamma ray flux into 4 latitude and 45 longitude regions the background and the DM annihilation signal were fitted to the data for each of the 180 bins.
The DM annihilation flux is dominated by the annihilation flux of the clumpy DM which is proportional to $\rho_{\chi, clump}$, not $\rho^2_{\chi, clump}$. While the diffuse DM component has a cuspy NFW profile a shallower cored distribution was obtained for the clumpy component. The DM annihilation flux is dominated by the clumpy DM component, but the clumpy component yields a mass below the required mass $> 10^{12}$ solar masses \cite{RefBattaglia}. However, if combined with the diffuse cuspy NFW profile, both the EGRET data and the mass constraint can be fulfilled.
In order to take the tidal disruption of DM clumps in the vicinity of stars into account a survival probability for clumps was introduced. Most of the clumps are expected to be destroyed near the Galactic centre, although a steep cusp may survive. The strong signal observed from the Galactic centre yielded a survival probability at the centre of $P(0)=0.7$. This means that the DM clumps are not completely destroyed, which is in good agreement with more detailed calculations in Ref. \cite{RefDokuchaev2}. 
In addition to the DM halo profiles two ringlike substructure were required at radii of 4 and 13 kpc. The halo and ring parameters were obtained by minimizing a $\chi^2$ function comparing the flux of the excess from the various sky directions with the line-of-sight integral in the halo.
Figure \ref{fig:longitudes} shows that the halo model fits the measured data very well.\\
In summary, the EGRET excess of diffuse Galactic gamma rays is in good agreement with the expectations of a cored clumpy halo component plus a cuspy diffuse one. The ringlike substructure, expected from the tidal disruption of the nearby Canis Major dwarf galaxy, is clearly seen and its heavy mass above $10^{10}$ solar masses as obtained from the EGRET data, has been recently confirmed by the reduced gas flaring in this region.

\end{document}